\documentclass[final
  ]
  {aipproc}

\layoutstyle{6x9}

\usepackage{graphicx}
\usepackage{epsfig}
\usepackage{latexsym}
\newcommand{\ba}{\begin{array}}
\newcommand{\ea}{\end{array}}

\newcommand\lsim{\mathrel{\rlap{\lower4pt\hbox{\hskip1pt$\sim$}}
    \raise1pt\hbox{$<$}}}
\newcommand\gsim{\mathrel{\rlap{\lower4pt\hbox{\hskip1pt$\sim$}}
    \raise1pt\hbox{$>$}}}

\newcommand{\nn}{\nonumber \\}

\newcommand{\mrm}{\mathrm}

\newcommand{\Sres}{{\cal S}}
\newcommand{\Pres}{{\cal P}}
\newcommand{\model}{{\Sres\Pres}}

\newcommand{\Cm}{C^\model}
\newcommand{\MS}{M_S}
\newcommand{\MP}{M_P}

\newcommand{\mbf}{\mathbf} 

\begin{document}

\title{Progress in Chiral Perturbation Theory}

\classification{11.30.Rd,12.39.Fe,12.15.Hh,14.60.Ef}
\keywords      {Effective field theory, chiral perturbation theory,
  $V_{us}$, $(g-2)_\mu$}

\author{Gerhard Ecker}{
  address={Inst. Theor. Physik, Univ. Wien, Boltzmanng. 5, A-1090
  Wien, Austria}
}

\begin{abstract}
After a short status report on chiral perturbation theory, I review
recent progress in determining some of the low-energy couplings by
matching the effective theory to QCD. Consequences for $K_{l3}$ decays 
and for the extraction of the CKM matrix element $V_{us}$
are reported. Hadronic vacuum
polarization at low energies and its impact on the anomalous magnetic
moment of the muon are discussed. 
\end{abstract}

\maketitle


\section{Introduction}
At a time when the LHC is getting ready to open a new era in particle
physics, it is legitimate to ask why one should still be interested in
QCD (more generally in the Standard Model) at low energies. There
are at least two good reasons to pursue the study of QCD in the
confinement regime.
\begin{itemize} 
\item It is a challenge for theoretical particle physics to derive
  reliable results in the nonperturbative domain. An impressive example
  is pion-pion scattering, one of the few examples in hadron
  phenomenology at low energies where theory is ahead of experiment
  \cite{Colangelo:2001df}. Important information on the mechanism of
  spontaneous chiral symmetry breaking can be extracted from pion-pion
  scattering, especially from S-wave scattering lengths. The study of
  QCD in the nonperturbative regime may turn out
  to be relevant even for LHC physics if the simple Higgs mechanism of
  the Standard Model turns out to be insufficient to describe
  electroweak symmetry breaking.  
\item The assessment of physics beyond the Standard Model will remain
  an important research topic even 
  at much lower than LHC energies. The reduction in energy
  must be compensated by an increase in precision, both in experiment
  and in theory. In the long run, lattice gauge theories and 
  low-energy effective field theories will survive as the most 
  comprehensive and reliable approaches in this field.
\end{itemize} 
In this talk, I present a short progress report on chiral
perturbation theory (CHPT), which is precisely the effective field
theory of the Standard Model at low energies. 
Green functions and amplitudes are dominated 
at low energies by the exchange of pseudoscalar mesons,
the pseudo-Goldstone bosons of spontaneously broken chiral symmetry,
allowing for a systematic expansion in momenta and quark masses. I
discuss recent progress in determining some of the a priori unknown
coupling constants of CHPT, a recent CHPT analysis of $K_{l3}$
decays to extract the CKM matrix element $V_{us}$ and,
finally, very recent developments concerning the determination of
hadronic vacuum polarization, a
topic of great importance for comparing the Standard Model prediction of
the muon magnetic moment with experiment. The extension of CHPT to
the intermediate-energy region dominated by meson resonances is
covered by J. Portol\'es \cite{JP_QCDW}.

\section{Status of chiral perturbation theory}

The spontaneously and explicitly broken chiral symmetry of QCD is the
key feature of CHPT. The corresponding Lagrangian is organized in an
expansion in derivatives (vestige of spontaneous symmetry breaking)
and in quark masses (explicit breaking). CHPT is a nonrenormalizable
quantum field theory that must nevertheless be renormalized like any
respectable quantum field theory. The main difference to
renormalizable theories is the rapidly increasing number of low-energy
constants (LECs) in higher orders of CHPT. As a low-energy effective
field theory, CHPT can be applied to processes with momenta $\ll$ 1
GeV. 

In the mesonic sector, the
original effective chiral Lagrangian of next-to-leading order 
\cite{Gasser:1983yg,Gasser:1984gg} has been extended to
next-to-next-to-leading order \cite{Bijnens:1999sh} or
$O(p^6)$ in the standard chiral counting. At this order, diagrams with
up to two loops have to be taken into account for a consistent low-energy
expansion (see Ref.~\cite{Bijnens:2004pk} for a recent review).  

Still in the meson sector, the formalism of CHPT has been extended to
incorporate the nonleptonic weak interactions and to implement
radiative corrections for strong processes as well as for semileptonic 
and nonleptonic weak decays. The corresponding Lagrangians and the 
associated number of LECs are displayed in Table \ref{tab:EFTSM}. As
the Table indicates, the state of the art for these extensions is
next-to-leading order with at most one-loop amplitudes.

\renewcommand{\arraystretch}{1.3}
\begin{table}[h!]
\caption{The effective chiral Lagrangian of the SM in the
  meson sector. The numbers in brackets refer to the number of
independent couplings for $N_f=3$. 
The parameter-free Wess-Zumino-Witten action 
  that cannot be written as the four-dimensional integral of an
  invariant Lagrangian must be added.}
\label{tab:EFTSM}
\vspace{.5cm}
\begin{tabular}{|l|c|} 
\hline
&  \\
\hspace{1cm} ${\cal L}_{\rm chiral\; order}$ 
~($\#$ of LECs)  &  loop  ~order \\[8pt] 
\hline 
&  \\
${\cal L}_{p^2}(2)$
~+~${\cal L}_{G_Fp^2}^{\Delta S=1}(2)$  
~+~${\cal L}_{e^2p^0}^{\rm em}(1)$
~+~${\cal L}_{G_8e^2p^0}^{\rm emweak}(1)$ & $L=0$ \\[15pt]
~+~${\cal L}_{p^4}(10)$~+~${\cal L}_{p^6}^{\rm odd}(32)$
~+~${\cal L}_{G_8p^4}^{\Delta S=1}(22)$
~+~${\cal L}_{G_{27}p^4}^{\Delta S=1}(28)$ &   
$L=1$ \\[3pt]
~+~${\cal L}_{e^2p^2}^{\rm em}(14)$
~+~${\cal L}_{G_8e^2p^2}^{\rm emweak}(14)$ 
~+~${\cal L}_{e^2p}^{\rm leptons}(5)$  & \\[15pt] 
~+~${\cal L}_{p^6}(90)$  & $L=2$ \\[8pt] 
\hline
\end{tabular}
\end{table}

Effective chiral Lagrangians have also been employed for baryonic
processes \cite{Bernard:1995dp} and for light nuclei
\cite{Bedaque:2002mn}.

\section{Low-energy constants}
As Table \ref{tab:EFTSM} shows, a major problem of CHPT is the
abundance of LECs in higher orders of the chiral expansion. For a
phenomenological determination of those constants, two types of LECs
can be distinguished.
\begin{itemize} 
\item[i.] The associated contributions survive in the chiral limit. Such
  LECs govern the momentum dependence of amplitudes and are at least
  in principle accessible experimentally.
\item[ii.] The couplings are associated with explicit chiral symmetry
  breaking. Such LECs specify the quark mass dependence of
  amplitudes. They are difficult if not impossible to extract from
  experiment but they are accessible in lattice QCD.
\end{itemize}
However, at the present level of sophistication it is unrealistic to
expect a phenomenological determination of all LECs even of type i
only. Instead, some progress has been made recently in matching CHPT
to QCD by investigating specific Green functions in the limit of large
$N_C$. As in every effective field theory, the
LECs are sensitive to the ``heavy'' degrees of freedom not represented
by explicit fields in the Lagrangian. Experience shows that 
truncation of the infinitely many intermediate states (for $N_C \to
\infty$) to the lowest-lying resonances is usually sufficient.

Instead of reviewing the matching procedure in general, I discuss two
specific examples recently considered that have some impact on topics
of current interest.

\subsection{Radiative semileptonic decays}
In the discussion of radiative corrections for semileptonic kaon
decays the Lagrangian ${\cal L}_{e^2p}^{\rm leptons}$
\cite{Knecht:1999ag} in Table \ref{tab:EFTSM} enters. In a two-step 
procedure, the Fermi theory of semileptonic decays was matched to both 
the Standard Model and CHPT \cite{Descotes-Genon:2005pw} resulting in 
spectral representations for all five LECs in 
${\cal L}_{e^2p}^{\rm leptons}$. 

Let me concentrate here on one of those LECs ($X_1$) that will be
relevant later on. The authors of Ref.~\cite{Descotes-Genon:2005pw} 
obtain the following representation for $X_1$,
\begin{equation} 
X_1 = \displaystyle\frac{3i}{8} \displaystyle\int
\displaystyle\frac{d^4 k}{(2\pi)^4}\left(\Gamma_{VV}(k^2) -
\Gamma_{AA}(k^2)  \right)/k^2 ~,
\end{equation}
in terms of vertex functions ($V_\mu^a$ is an $SU(3)$ vector current
and $\phi^c$ is a member of the pseudoscalar octet)
\begin{equation}
\Gamma_{VV}(k^2) \sim \displaystyle\lim_{p \to 0} 
\displaystyle\int d^4 x e^{i k x} \langle 0|T\, V_\mu^a(x) V_\nu^b(0)|
\phi^c(p) \rangle
\end{equation} 
and similarly for $\Gamma_{AA}(k^2)$. The integral converges well and,
when saturated with the lowest-lying $V,A$ meson resonances, produces
a value $X_1=-0.0037$ \cite{Descotes-Genon:2005pw} to be used for the
analysis of $K_{l3}$ decays.  

\subsection{Strong LECs of $\mbf{O(p^6)}$}
The second example concerns LECs that appear in the $K_{l3}$
amplitudes at $O(p^6)$. The Green function of interest is the
three-point function of scalar and pseudoscalar densities:
\begin{equation} 
i^2\! \int \! dx\,  dy \, e^{i px +i qy + irz} \langle 0 | T S^a (x) P^b
(y) P^c (z) | 0 \rangle =  d^{abc} \,\Pi_{SPP} (p^2,q^2,r^2) ~.
\end{equation}  
At low energies, $\Pi_{SPP}$ is given in terms of LECs of 
$O(p^4)$ and $O(p^6)$ since loop contributions are subdominant for
large $N_C$. At high momenta, the operator product expansion (OPE)
fixes the behaviour of $\Pi_{SPP}$ that vanishes in QCD perturbation 
theory as an order parameter of spontaneous chiral symmetry breaking. 
Additional constraints apply for (transition) form factors at
large momentum transfer, with two external momenta on shell.

To interpolate between CHPT and QCD, a large-$N_C$ motivated ansatz
can be employed \cite{Cirigliano:2005xn}:
\begin{equation}
\Pi_{SPP}^\model(s,t,u) =
\frac{P_0+P_1+P_2+P_3+P_4}{[\MS^2 -s][-t][-u][\MP^2-t][\MP^2-u]}~,
\end{equation}
with polynomials $P_n$ of degree $n$ in $s,t,u$ (altogether 21
parameters). The OPE limits $n \le 4$ and
lowest-order CHPT fixes the constant $P_0$. The high-energy conditions
constrain the polynomials $P_1, P_2$ of direct relevance for the
LECs. The final relations for the $O(p^6)$ LECs of interest are
\cite{Cirigliano:2005xn}
\begin{eqnarray} 
\Cm_{12} = - \displaystyle\frac{F^2}{8
\MS^4}~,\, &
\Cm_{34} =   \displaystyle\frac{3 \, F^2}{16 \MS^4} + 
\displaystyle\frac{d_m^2}{2} 
\left(\displaystyle\frac{1}{\MS^2} - 
\displaystyle\frac{1}{\MP^2} \right)^2
\end{eqnarray} 
in terms of the masses $M_S, M_P$ of the lowest-lying (pseudo-)scalar
nonets, the pion decay constant $F$ and a resonance coupling 
$d_m ~\sim ~F/(2\sqrt{2})$. All parameters refer to the chiral limit.

The first interpretation of these results is not too encouraging. 
There are big uncertainties related to the value of $M_S$ in
particular and to the rather strong scale dependence of $C_{12}$ and 
$C_{34}$, which is however inaccessible at leading order in $1/N_C$.

\section{$\mbf{K_{\lowercase{l3}}}$ and $\mbf{V_{\lowercase{us}}}$}

The analysis of $K_{l3}$ decays allows for the presently most accurate
determination of the CKM matrix element $V_{us}$. In general, two form
factors characterize the decay matrix element:
\begin{equation} 
\langle \pi^- (p_\pi) | \bar{s} \gamma_\mu u | K^0 (p_K) \rangle = 
f_{+}^{K^0 \pi^-} (t)  \, (p_K + p_\pi)_\mu  + 
f_{-}^{K^0 \pi^-} (t)  \, (p_K - p_\pi)_\mu ~.
\end{equation}
Of special interest for the determination of  $V_{us}$ is the quantity
$f_{+}^{K^0 \pi^-} (0)$ with the following chiral expansion:
\begin{equation} 
f_{+}^{K^0 \pi^-} (0) = 1 + f_{p^4} + f_{e^2\,p^2} + f_{p^6} + O[(m_u
- m_d)p^4,e^2\,p^4] ~.
\end{equation}
The present status is as follows:
\begin{center}  
\begin{tabular}{lclcl} 
$f_{p^4}$ & \mbox{}\hspace*{1cm}   & $- 0.0227$ 
(no uncertainty) & \mbox{}\hspace*{1cm} & \cite{Gasser:1984ux} \\
$f_{e^2\,p^2}$ & & radiative corrections ($X_i$) &
  \mbox{}\hspace*{1cm} & \cite{Cirigliano:2004pv} \\ 
$f_{p^6}$ & & loop contributions & \mbox{}\hspace*{1cm} & 
\cite{Post:2001si,Bijnens:2003uy} \\
& & tree contributions & \mbox{}\hspace*{1cm}  & $L_5^2$, $C_{12}+C_{34}$
\end{tabular}
\end{center} 
For a first comparison with experiment, consider the ratio
\cite{Cirigliano:2004pv}
\begin{eqnarray}
r_{+0} := \left( \frac{2 \, \Gamma(K^+_{e 3
(\gamma)}) \, M_{K^0}^5 \, I_{K^0}}{\Gamma(K^0_{e 3 (\gamma)}) \,
M_{K^+}^5 \, I_{K^+}} \right)^{1/2} = 
\displaystyle\frac{|f_{+}^{K^+ \pi^0} (0)|}{|f_{+}^{K^0 \pi^-} (0)|}
~. 
\end{eqnarray} 
The theoretical prediction for $r_{+0}$ is independent of
$f_{p^6}$. The only previously unknown LEC in $r_{+0}$ is $X_1$. With
the newly determined value for $X_1$ 
\cite{Descotes-Genon:2005pw} and using quadratic fits for the form
factors to extract $f_+(0)$ from the data, one finds 
\cite{Mescia:2004xd,HN05}
\begin{eqnarray}
r_{+0}^{\rm th} &=& 1.023 \pm 0.003 \nn
r_{+0}^{\rm exp} &=&  1.036 \pm 0.008 ~.
\end{eqnarray}
A possible discrepancy between theory and experiment for $r_{+0}$
could be due to several reasons:~radiative corrections applied by
experimentalists are not always state of the art, the lifetimes of
$K^+, K_L$ may still undergo revisions  and the error in 
$r_{+0}^{\rm th}$ due to neglected effects of 
$O[(m_u - m_d)p^4,e^2\,p^4]$ could be underestimated. 

\begin{figure}[!t]
\centering
\begin{picture}(300,175)  
\put(110,65){\makebox(50,50){\epsfig{figure=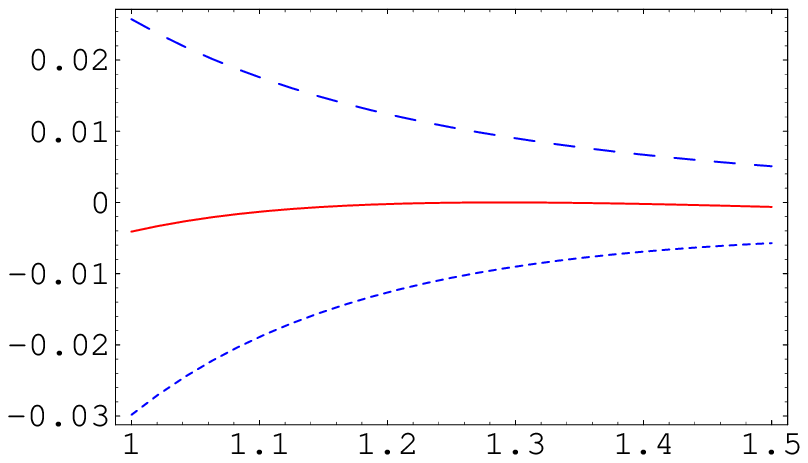,width=10cm}}}
\put(260,-5){
$M_S \ ({\rm GeV})$ 
}
\put(-55,160){
$ f_{p^6}^{\rm tree} (M_\rho)  $ 
}
\put(110,140){
{\small
$L_5 \times L_5 / F_\pi^2 $
}
}
\put(100,50){
{\small
$- (C_{12} + C_{34})$
}
}
\end{picture}
\caption{
$f_{p^6}^{\rm tree} (M_\rho)$ is displayed as a function of $M_S$ 
for $M_P=1.3$ GeV (solid line). The dashed line represents  the 
term proportional to $L_5^2$, while the 
dotted line represents the term proportional to 
$- (C_{12} + C_{34})$.
}
\label{fig:spp}
\end{figure}

Turning now to  $f_{+}^{K^0 \pi^-} (0)$,
the uncertainty in the $O(p^6)$
contribution $f_{p^6}$ is mainly due to the LECs. Loop and local
contributions are separately scale dependent. The loop contributions
at the scale $\mu=M_\rho$ amount to \cite{Bijnens:2003uy}
\begin{equation} 
f_{p^6}^{L=1,2} (M_\rho) =  0.0093 \pm 0.0005 ~.
\end{equation}
The local contribution is given by
\begin{equation}
f_{p^6}^{\rm tree} (M_\rho) =
8 \frac{\left( M_K^2 - M_\pi^2 \right)^2}{F_\pi^2}  
\, \left[\frac{\left(L_5^r (M_\rho) \right)^2}{F_\pi^2} - 
C_{12}^r (M_\rho) - C_{34}^r (M_\rho) \right] ~.
\end{equation}
The results of large-$N_C$ matching discussed in the previous section
can be read off from Fig.~\ref{fig:spp}. The separate contributions
$L_5^2$ and $C_{12} + C_{34}$ depend strongly both on the uncertain
scalar resonance mass $M_S$ and on the
renormalization scale. However, as shown in Fig.~\ref{fig:spp} for the
$M_S$ dependence, both uncertainties are substantially reduced
for the relevant combination entering  $f_{p^6}^{\rm tree} (M_\rho)$. 

A strong destructive interference between the two local contributions
is observed. The final result (allowing for a second pseudoscalar
multiplet $P^\prime$) is \cite{Cirigliano:2005xn}
\begin{eqnarray}
f_{p^6}^{\rm tree} (M_\rho) &=& - 0.002  \pm 0.008_{\, 1/N_C} \pm 
0.002_{\, M_S} \,\mbox{}_{- 0.002}^{+0.000} \,\mbox{}_{\, P^\prime} \nn
f_{p^6} &=& 0.007 \pm 0.012  \nn
f_{+}^{K^0 \pi^-} (0) &=&  0.984 \pm 0.012 ~.
\end{eqnarray}
We find less $SU(3)$ breaking in $f_{+}^{K^0 \pi^-} (0)$ compared to
Leutwyler and Roos \cite{Leutwyler:1984je}, with $f_{p^6}$ being 
dominated by the loop contribution. From the experimental result 
\cite{Mescia:2004xd} $f_{+}^{K^0 \pi^-} (0) \cdot |V_{us}| =
0.2160(10)$ one obtains 
\begin{equation}
|V_{us}| = 0.2195 \pm 0.0027_{f_+(0)} \pm 0.0010_{\rm exp} ~.
\end{equation}
Before observing a possible conflict with CKM unitarity (the PDG value
\cite{pdg04} for $V_{ud}$ gives rise to $|V_{us}|^{\rm unitarity} =
0.2265 \pm 0.0022$), the following remarks are in order.
\begin{itemize} 
\item[i.] A new result for the neutron lifetime \cite{Serebrov:2004zf}
  would prefer a value for $V_{ud}$ in perfect agreement with 
  $|V_{us}| = 0.2195$ and unitarity.
\item[ii.] A recent analysis of semileptonic hyperon decays 
  \cite{Acosta:2005iw} yields $|V_{us}| = 0.2199 \pm 0.0026$. After
  the Workshop, the uncertainties of extracting $V_{us}$ from
  semileptonic hyperon decays have been reassessed in 
  Ref.~\cite{Mateu:2005wi}. 
\item[iii.] To achieve an accuracy of better than 1\% for $V_{us}$, the 
  differences between $K^+$ and $K^0$ results must be straightened
  out. 
\end{itemize}  
An independent check of the theoretical estimate for the LECs of
$O(p^6)$ is provided by the slope $\lambda_0$ of the scalar form 
factor (accessible in $K_{\mu 3}$ decays) that depends on the same 
LECs $C_{12},C_{34}$ as $f_{+}(0)$. 
\begin{center} 
\begin{tabular}{ccccc} 
Ref. & \mbox{} \hspace*{1cm} & Cirigliano et al. 
\cite{Cirigliano:2005xn}  & \mbox{} \hspace*{1cm} &
  KTeV \cite{Alexopoulos:2004sy} \\[.2cm]  
$\lambda_0\cdot 10^3$ & & $13 \pm 3$ &  & $13.72 \pm 1.31$
\end{tabular} 
\end{center} 

\section{Hadronic vacuum polarization and $\mbf{(\lowercase{g}-2)_\mu}$}

\begin{figure}[ht]
  \includegraphics[height=.2\textheight]{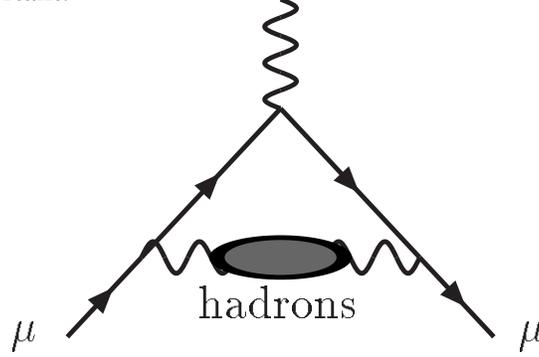}
  \caption{Contribution of lowest-order hadronic vacuum polarization
  to the muon magnetic moment.}
\label{fig:amu}
\end{figure}

At present, the biggest uncertainty in the evaluation of the anomalous
magnetic moment of the muon $a_\mu$ in the Standard Model is due to 
hadronic vacuum polarization at lowest order in $\alpha$ (shown in
Fig.~\ref{fig:amu}) that is directly
related to the cross section $\sigma(e^+ e^- \to$ hadrons). About 73
\% of $a_\mu^\mrm{vac.pol.}$ comes from the $\pi^+ \pi^-$ final state,
the low-energy part being especially important. 
 
In the isospin limit, the two-pion contribution to hadronic vacuum 
polarization can also be obtained from the decay $\tau^- \to \pi^-  
\pi^0 \nu_\tau$ \cite{Alemany:1997tn}. At the level of accuracy needed
for a comparison with the measured value of $a_\mu$
\cite{Bennett:2004pv}, isospin violating and electromagnetic
corrections must be included \cite{Cirigliano:2002pv,Davier:2003pw}. 

However, until recently the two-pion spectral functions from $e^+ e^-$
annihilation and from $\tau$ decays seemed to differ significantly
especially above the $\rho$ region, even after accounting for isospin 
violating effects. The value of $a_\mu^{\pi\pi}$ on the basis of the 
most precise $e^+ e^-$ data from the CMD-2 Collaboration
\cite{Akhmetshin:2003zn}  was then confirmed by KLOE 
\cite{Denig:2005eb} although the actually measured $\pi\pi$ cross
sections are not in very good agreement. The consensus among many
experts in the field was spelled out by H\"ocker at last year's
High Energy Conference in Beijing \cite{Hocker:2004xc}: until the
origin  of the discrepancy between $e^+ e^-$ and $\tau$ data is 
understood the $\tau$ data should be ignored for the evaluation of 
$a_\mu$. 

A recent analysis of Maltman \cite{Maltman:2005yk} suggests a new 
perspective on this issue. He investigates so-called pinched FESR of
the type
\begin{equation} 
\displaystyle\int_0^{s_0} w(s) \rho(s) ds =
-\displaystyle\frac{1}{2\pi\,i} \oint_{|s|=s_0} w(s) \Pi(s) ds 
\label{FESR} 
\end{equation}
for current correlators $\Pi(s)$ with associated spectral functions
$\rho(s)$. The spectral functions of interest here are the
electromagnetic spectral function $\rho_{\rm em}$ measured in $e^+
e^-$ annihilation and the charged $I=1$ vector current spectral
function $\rho_V^{I=1}$ accessible in $\tau$ decays. The weight
function $w(s)$ is a positive definite analytic function in the
complex $s$-plane for $|s| \le s_0$, but otherwise arbitrary except
for the constraint $w(s_0)=0$ to minimize duality violations
(pinching). 

The left-hand side of the FESR (\ref{FESR}) is evaluated with
experimental input (CMD-2 \cite{Akhmetshin:2003zn} and ALEPH 
\cite{Barate:1997hv}) whereas the right-hand side is calculated from 
QCD with the help of the OPE. The freedom
of choosing the weight function $w(s)$ can be employed to eliminate
the dimension $D=6$ OPE contributions altogether. The right-hand side
is then mainly sensitive to the $D=0$ perturbative part known up to
$O(\alpha_s^3)$, with weaker
dependences on $m_s$ (in the $D=2$ piece) and on $D=4$ quark and gluon
condensates. Effects with $D \ge 8$ can be kept under control by
varying $s_0$. Discarding all low-energy input for the determination
of $\alpha_s(M_Z)$ (such as the $\tau$ data that are to be tested with
FESR), Maltman obtains a value
\begin{equation} 
\alpha_s(M_Z)=0.1200 \pm 0.0020
\label{as}
\end{equation} 
to be used for the right-hand side of (\ref{FESR}). 

A first test
performed in Ref.~\cite{Maltman:2005yk} consists in fitting
$\alpha_s(M_Z)$ from the experimentally determined spectral integrals
(left-hand side), leaving all other input for the right-hand side
unchanged. The results for two typical weight functions $w_1, w_6$ are
shown in Table \ref{tab:as}, to be compared with the best value from
high-energy data in Eq.~(\ref{as}).
Taking into account that the weights are positive definite, the
results in Table \ref{tab:as} indicate that the electromagnetic
spectral density is too low whereas the $\tau$ spectral data are in
perfect agreement with the canonical value of $\alpha_s$. 
\begin{center} 
\begin{table}[h!]
\label{tab:as}
\caption{Fitted values of $\alpha_s(M_Z)$ from experimentally
  determined spectral integrals for two different weight
  functions \cite{Maltman:2005yk}. }
\begin{tabular}{|c|c|c|} 
\hline
weight & \hspace*{.2cm} type \hspace*{.2cm}  & $\alpha_s(M_Z)$
 \\
\hline
$w_1$ & em & 0.1138$\begin{array}{l} + 0.0030 \\
                                   - 0.0035  \end{array}$ \\
$w_6$ & em & 0.1150$\begin{array}{l} + 0.0022 \\
                                   - 0.0026  \end{array}$ \\
$w_1$ & $\tau$ & 0.1218$\begin{array}{l} + 0.0027 \\
                                   - 0.0032  \end{array}$ \\
$w_6$ & $\tau$ & 0.1201$\begin{array}{l} + 0.0020 \\
                                   - 0.0022  \end{array}$ \\
\hline
\end{tabular}
\end{table}
\end{center} 
A second independent consistency check of the data comes from a 
comparison of the two sides in Eq.~(\ref{FESR}) for different values of 
$s_0$ \cite{Maltman:2005yk}. When plotting the spectral integrals as
functions of $s_0$ one arrives at a similar conclusion as before: the
slopes in the electromagnetic case differ by about 2.5 $\sigma$
between data and QCD. On the other hand, the $\tau$ data show perfect
consistency both for the slope and in absolute normalization
(depending on $\alpha_s$).

The conclusions of Ref.~\cite{Maltman:2005yk} are very convincing even
if the statistical weight is not overwhelming: the sum rule tests
clearly favour the $\tau$ over the $e^+ e^-$ data. The status of
$a_\mu$ at the time of the Workshop can be summarized as follows 
\cite{Davier:2004gb}:
\begin{equation}
\left(a_\mu^\mrm{exp} - a_\mu^\mrm{SM}\right)\cdot 10^{10} 
=\left\{\begin{array}{ll}
23.9 \pm 9.9 & (2.4 ~\sigma) ~~[e^+ e^-]\\
7.6 \pm 8.9 & (0.9 ~\sigma) ~~[\tau, e^+ e^-]
\end{array} \right. ~.
\end{equation}
Using the isospin corrected $\tau$ data for the $2\pi$ and $4\pi$
final states thus leads to agreement between theory and experiment to 
better than 1 $\sigma$.

Two weeks after the Workshop, new $e^+ e^- \to \pi^+ \pi^-$ data were
released \cite{Achasov:2005rg} that appear to lie between the CMD-2
and the (isospin corrected) ALEPH data.

\section{Conclusions}
In the meson sector, chiral perturbation theory has been pushed to
next-to-next-to-leading order. At this order, the main limitation for
further progress is the abundance of coupling constants, an
unavoidable feature of a nonrenormalizable effective field
theory. Some progress has been made recently in estimating those
constants by using large-$N_C$ methods to interpolate between CHPT 
and QCD.

CHPT is the only reliable approach for calculating electromagnetic and
isospin violating corrections for hadronic processes at low
energies. This is in particular important for the analysis of $K_{l3}$
decays in order to extract the CKM matrix element $V_{us}$ to better
than 1 \% accuracy.

Recent sum rule tests \cite{Maltman:2005yk} favour $\tau$ over $e^+
e^-$ data for evaluating the hadronic vacuum polarization at low and
intermediate energies. As a
consequence, there is at present no conflict between the Standard
Model and experiment for the anomalous magnetic moment of the muon.


\begin{theacknowledgments}
This work has been supported in part by HPRN-CT2002-00311
(EURIDICE). I thank Helmut Neufeld for discussions and for numerical 
help with the ratio $r_{+0}$. Special thanks and congratulations to 
Pietro Colangelo, Fulvia De Fazio, Giuseppe Nardulli and their team 
for a very successful Workshop.
\end{theacknowledgments}

\end{document}